\documentclass{article}
\usepackage{amsmath, amssymb, graphics, setspace}

\newcommand{\mathsym}[1]{{}}
\newcommand{\unicode}[1]{{}}

\begin{document}

\title{Strain mediated adatom stripe morphologies on Cu$<$111$>$ simulated.}
\author{Wolfgang Kappus\\
wolfgang.kappus@t-online.de}
\date{v02: 2013-01-29}
\maketitle

\section*{Abstract}

Substrate strain mediated adatom configurations on Cu$<$111$>$ surfaces have been simulated { }in a coverage range up to nearly 1 monolayer. Interacting
adatoms occupy positions on a triangular lattice in two dimensions. The elastic interaction is taken from earlier calculations, short range effects
are added for comparison. Dependent on the coverage different morphologies are observed: Superlattices of single adatoms in the 0.04 ML region, ordered
adatom clusters in the 0.1 ML region, elongated islands in the 0.3 ML region, and interwoven stripes in the 0.5 ML region. In the region above the
sequence is reversed with occupied and empty positions complemented. Stronger short range interactions increase the feature size of the clusters
and reduce their lattice order. The influence of the substrate elastic anisotropy turns out to be significant. Results are compared with morphologies
observed on Cu$<$111$>$ surfaces and the applicability of the model is discussed.

\section*{1.Introduction}

Regular self-assembled adatom structures, ranging from superlattices via nanodot arrays to strain relief pattern are interesting for various general
and technological reasons, reviews were given in [1,2]. While interactions of adatoms comprise various mechanisms [3] the focus on elastic interactions
in this paper is driven by the question on their importance compared with other interactions. In recent calculations on the stability and dynamics
of strain mediated superlattices it was shown that the role of elastic interactions was underestimated compared with surface state mediated interactions
[4], so other surface phenomena seem worth to be discussed in the light of strain mediated interactions. Anisotropies of adatom pattern can act as
a probe for indicating strain mediated interactions via correlation to anisotropies of the substrate elastic constants.

The calculations on the stability and dynamics of strain mediated superlattices [4] covered a low coverage region and left the question open how
adatoms arrange under equilibrium conditions when the coverage is increased. The experiments of Plass et al. on domain patterns [5] provide a challenge
to prove the ability of an elastic continuum theory in building a bridge between superlattices and stress relief patterns. Such bridge was built
before with a Green function method for the non-equilibrium case [6]. { }

The focus on Cu$<$111$>$ has good reasons as well: Cu is among the substrates with the highest elastic anisotropies and Cu$<$111$>$ seems to be a
preferred surface for experiments. Unfortunately the crystal directions are often not published, so a solid proof of elastic effects is hindered.
The predictions of this work are intended to allow a verification of the theory by experiments. { } 

The model used in this work to simulate adatom morphologies is an adaptation of the one used in [4]. The latter used a grid-less molecular dynamic
algorithm suited for low coverages. For the higher coverages up to 1 ML and the $<$111$>$ surface discussed in this work it had to be converted to
a grid base where adatom positions reside on a triangular lattice representing identical threefold coordinated substrate sites. The interaction mechanism
has been kept; it is based on the isotropic stress individual adatoms on threefold coordinated sites exert to their neighborhood. The limitations
of such interaction mechanisms and of other model assumptions are discussed below. { } 

The model results will be presented as sample adatom configurations for increasing coverages and for three variants of the interaction. The variants
stand for three different strengths of short range interactions and should give an idea of the interplay between short- and medium-term interactions.
The model results will also be presented as pair distributions derived from averaging over sample configurations. 

This work is organized as follows: In section 2 the details of the interaction model are recalled and the simulation model is detailed. Furthermore
the calculation method for pair distributions is described. In section 3 the model results will be presented as sample adatom configurations and
as pair distributions derived from averaging over sample configurations. For symmetry reasons the pair distributions will cover 30${}^{\circ}$ segments
only. In section 4 the model assumptions are reviewed, the model results are summarized and compared with a few experiments and open questions are
addressed. Section 5 closes with a summary of the results.

\section*{2.Model details}

In this section the elastic interactions used within the model are recalled, the grid based algorithm for the Molecular Dynamic simulations is introduced
and the method for deriving adatom pair distributions is explained. Also scaling relations, intended for the interpretation of experiments, are recalled.
{ } { }

\subsection*{2.1.Elastic interactions of adatoms }

Following [8] the interaction of adatoms located at the origin and at \(\overset{\rightharpoonup }{s}\) using polar coordinates (s,$\phi $) for their
distance s = $|$\(\overset{\rightharpoonup }{s}\)$|$ and pair direction angle $\phi $ with respect to the crystal axes is given by 

\[U(s,\phi ) =(2\pi )^{-1}\sum _p \omega _p \frac{\left.\cos (p \phi )\cos (p \frac{\pi }{2}\right)\Gamma \left(\frac{p+3}{2}\right)s^p\, _1F_1\left(\frac{p+3}{2};p+1;\frac{-s^2}{4
\alpha ^2}\right)}{2^{p+1}\Gamma (p + 1)\alpha ^{p+3}},\text{         }(2.1)\]

where \(\, _1F_1\) denotes the Hypergeometric Function, \(\Gamma (p)\) the Gamma function, $\alpha $=\(\sqrt{2}\)/2 is a cutoff length defining height
and location of the potential wall and the medium range potential, and the \(\omega _p\) denote coefficients of a cosine series describing the solution
of an elastic eigenvalue problem [8]. The dominating isotropic p=0 term of Eq. (2.3) is negative for small s describing a potential well (i.e. an
attractive potential), has a positive wall (i.e. a repulsive potential) at s=\(s_{w }\) and approaches infinity with a \(s^{-3}\) law. For elastic
anisotropic substrates like Cu { }the p$>$0 terms describe the anisotropic part of the interaction and influence the height of the positive wall
in dependence of the pair direction angle $\phi $ with respect to the crystal axes. Tab. 1 shows the \(\omega _p\) for the elastic adatom interaction
on Cu$<$111$>$ and W$<$111$>$ (for comparison) calculated as outlined in [8]. We note the units of the \(\omega _p\):\\
- the numerator is \(P^2\), the square of a scalar parameter P describing the lateral stress magnitude an adatom exerts to the surface\\
- the denominator is the \(c_{44}\) elastic constant of the substrate.\\
For details of the parameter P see [8]. \\

\begin{doublespace}
\noindent\(\pmb{
\begin{array}{|cccccccc|}
\hline
 \text{Substrate} & c_{11} & c_{12} & c_{44} & \zeta  & \omega _0 & \omega _6 & \omega _{12} \\
 \text{Cu} & 169. & 122. & 75.3 & -1.376 & -1.01 & -0.007 & +0.0004 \\
 W & 523. & 203. & 160. & 0. & -0.720 & 0. & 0. \\
\hline
\end{array}
}\)
\end{doublespace}

Table 1. Substrate Elastic Constants \(c_{\text{ik}}\) (GPa) from [9], anisotropy $\zeta $=(\(c_{11}\)-\(c_{12}\)-2\(c_{44}\))/\(c_{44}\) and coefficients
\(\omega _p\) (in \(P^2\)/\(c_{44}\) units) { }on Cu $<$111$>$ and W$<$111$>$ .$\quad $

In the present analysis the strong attractive interaction of Eq. (2.1) in the region s$<$\(s_{w }\) is replaced by three variants to study the influence
of short range interactions - in addition to the elastic interaction - on the medium range adatom morphology: \\
- variant 1 as used and described in [4]

\[U_1(s,\phi ) =U_w+U_{\text{wp}}\cos (\text{p$\phi $}) \frac{s}{s_w}\text{       }\text{for} s<s_w ,\text{      }(2.2)\]

where \(U_w\) describes the wall height, \(U_{\text{wp}}\) the wall anisotropy variance, and \(s_w\) is the location of the wall maximum,\\
- variant 2, describing additional attraction between next neighbors

\[U_2(s,\phi ) =0\text{       }\text{for} s<s_0 ,\text{       }(2.3)\]

where \(s_0\) is defined by \(U\)(\(s_0\),$\phi $) = 0, covering the range s$\lesssim $1.75, significantly smaller than \(s_w\),

- variant 3, describing stronger attraction between next neighbors

\[U_3(s,\phi ) =-5\text{  }k_B T\text{     }\text{for} s\leq s_3 ,\text{       }(2.4)\]

where \(s_3\) is the next neighbor distance. The value -5 is chosen to get an equidistant series of U values. 

\subsection*{2.2.Simulations }

The Molecular Dynamics grid-less algorithm used in [4] turned out unstable and inefficient in the coverage range $\theta >$0.1. Therefore a triangular
grid algorithm has been used instead. The triangular grid represents adatom positions on a $<$111$>$ surface with threefold symmetry fulfilling the
symmetry condition used for the adatom generated surface stress [8]. Periodic boundary conditions were applied to avoid the problem of adatom diffusion
to the boundary. The hexagon diameter of 48 units was chosen to keep the computing time in the range of hours while the interaction u(s=24) has decreased
well below 0.01. Temperature effects are treated by the normalized interaction

\[u(s,\phi )=U(s,\phi ) \left/k_B\right. T.\text{          }(2.5)\]

Not knowing the size of the stress parameter P and as in [4] the average wall height is assumed \(u_W\)=5 and this choice determines all u(s,$\phi
$).

In our grid algorithm an adatom configuration is described by a set of occupation numbers \(\left\{\tau _i\right\}, \tau _i\in \{0,1\}.\text{}\)Starting
from a random k member adatom configuration $\{$\(\tau _{i,0}\)\(\left\}, \text{step} n+1 \left\{\tau _{i,n+1}\right\} \text{evolves} \text{from}
\text{step} n \left\{\tau _{i,n}\right\} \text{by}\right.\) comparing the total interaction of each adatom i

\[u_{\text{tot}}(i)=\sum _{j=1}^k u_{\text{ij}} \tau _j\text{     }(2.6)\]

with that of its empty next neighbor positions. If a next neighbor position m has less total interaction, the adatom i jumps to that position m.
So adatoms move around under the force field of all neighbors until all interaction is minimized.

The iterations are terminated when either no more jumps occur or when loops of identical configurations are detected. { }

\subsection*{2.3.Pair distribution }

The adatom pair distribution \(g_{\text{ik}}\) is calculated by averaging occupation pairs

\[g_{\text{ik}}=<\tau _i \tau _k>\left/\theta ^2\right.\text{  },\text{                            }(2.7)\text{      }\]

where $\theta $ denotes the coverage. So \textup{ \(g_{\text{ik}}\)}=1 in a random configuration, \textup{ \(g_{\text{ik}}\)}$>$1 if the pair $\{$\(\tau
_i\)\(,\tau _k\)$\}$ occurs more likely and \textup{ \(g_{\text{ik}}\)}$<$1 if the pair $\{$\(\tau _i\)\(,\tau _k\)$\}$ occurs less likely. It is
the discrete variant of g(s,$\theta $) calculated in [7] with a 2-dimensional Born-Green-Ivon type integral equation. 

\subsection*{2.4.Pair distribution scaling}

For the discussion of experimental results in section 4.5 we will need to recall scaling properties of the continuous pair distribution g(s,$\theta
$) as outlined in [7]. In the long range isotropic limit the adatom-adatom interaction becomes 

\[\text{   }u(s) = u_0 s^{-3} + O\left( s^{-5}\right)\text{  }, s>>s_0\text{    }(2.8)\text{     }\]

and the pair distribution scales 

\[\text{   }g\left(s,u_0,\theta \right) = g\left(\text{$\tau $s},\tau ^3u_0,\tau ^{-2}\theta \right)\text{         }(2.9)\text{    }\]

with a scaling factor $\tau $. In other words the pair distribution has the same shape if simultaneously the length is doubled, the interaction is
eightfold and the coverage is reduced by a factor of four. We also note from Eq. (2.5) an eightfold normalized interaction u results if { }the interaction
U is kept constant and the temperature T is reduced by a factor of eight. We further note that doubling the stress parameter P increases the interaction
U by a factor 4.

\section*{3.Results}

The results are presented in pairs of figures, the first of which shows a sample adatom configuration in a hexagon area simulated according to section
2.2 and the second shows an equivalent pair distribution \textup{ \(g_{\text{ik}}\)} according to section 2.3 and averaged over configuration samples.
The presentation comprises varying coverages $\theta $ and interactions \(U_i\) (sections 3.3 to 3.5) to study the influence of short range interactions.
We note the different notation of interactions \(U_i\) and scaled interactions \(u_i\) according to Eq. (2.5).

The adatom pair distributions are shown as dots at lattice positions in a 30 degree sector for symmetry reasons. A color code with different colors/
darkness is used to mark pair distribution ranges: \\
- \textup{ \(g_{\text{ik}}\)}$\geq $1.5 black { }\\
- 1.5$>$\textup{ \(g_{\text{ik}}\)}$\geq $1.0 blue/ dark gray\\
- 1.0$>$\textup{ \(g_{\text{ik}}\)}$\geq $0.5 green/ medium gray\\
- \textup{ \(g_{\text{ik}}\)}$>$0.5 yellow/ light gray\\
- \textup{ \(g_{\text{ik}}\)}$\leq $0.5 white. 

Since the algorithm used is different from the one previously used [4] the results section starts with a reference. Unfortunately a fault in the
code of [4] was just detected: { }the \(\cos (p \pi /2)\) term in Eq. (2.1) was omitted and therefore the results in [4] were rotated 30${}^{\circ}$
compared to the current corrected version.

\subsection*{3.1.Reference configuration}

Fig. 2.a shows empty (yellow points) and occupied positions (red points) of a triangular lattice. The interaction used is \(U_1\) and described by
Eqs. (2.1) and (2.2). Fig. 1.a acts as reference to [4] with a coverage { }$\theta $=0.045 to demonstrate that the new algorithm leads to the same
sample results except a 30${}^{\circ}$ rotation (as stated before). A substrate aligned superlattice of adatoms and a few dimers with a lattice parameter
of 5 grid units shows up like in the reference.

Fig. 1.b. shows the adatom pair distribution in a 30 degree sector taken from a configuration average. Black points near \(s_{<1-21>}\)=3*\(\sqrt{3}\)
and 6*\(\sqrt{3}\) lattice spacings and at \(s_{<1-10>}\)=9 reflect the (not quite perfect) aligned monomer superlattice. We note a blue dot at 1
\(s_{<1-10>}\)=1 reflecting a small population of next neighbor sites.

\subsection*{3.2.Influence of substrate elastic (an-)isotropy}

Previous investigations showed a strong influence of the substrate elastic constants on the adatom pair distribution [7]. A triangular grid algorithm
could compromise such delicate matter. To prove the grid algorithm properly handling substrate isotropy, the elastic constants of tungsten were used
as reference (see Tab.1). Fig.2 shows the resulting pair distribution for a coverage { }$\theta $=0.045 in a 30 degree sector. It shows (irrespective
statistical variances) the characteristic rings at 5, 10, 15 substrate lattice spacings already discussed in [7]. { } { }

Isotropy could also be compromised by adatom multiples. Though there are almost no angular moments of circular clusters in the relevant distance
of 5 lattice constants, adatom straight tripoles would generate differences in the interaction of up to 18$\%$ (less repulsive orthogonal to the
axes). 

\subsection*{3.3.Adatom configurations for coverages between 0.1 and 1 monolayer}

To stay consistent with [4] we will use in this section the short range interaction \(U_1(s,\phi )\), Eq. (2.2) and thus the more repulsive variant
1. In 0.2 steps the coverage is increased in Figs. 3.a. to 3.h showing the effects of a subsequent population of the 2-dimensional triangular lattice
up to 0.9 monolayers and the corresponding pair distributions according to Eq. (2.7). 

Fig.3.a shows a sample configuration at coverage $\theta $=0.1 and Fig.3.b shows the equivalent pair distribution taken from a configuration average.
The black dots in Fig.3.b near \(s_{<1-21>}\)=3*\(\sqrt{3}\) and at \(s_{<1-10>}\)=8 reflect a superlattice with a superlattice constant of nearly
5 substrate lattice spacings consisting of monomers, dimers, trimers and a few 4-mers. The dark dot at \(s_{<1-10>}\)=1 reflects the high amount
of next neighbors.

Fig.3.c shows a sample configuration at coverage $\theta $=0.3 and Fig.3.d shows the equivalent pair distribution taken from a configuration average.
The black dot in Fig.3.d near \(s_{<1-21>}\)=3*\(\sqrt{3}\) again reflects a superlattice with a superlattice constant of nearly 5 substrate lattice
spacings consisting of circular and elongated n-mers. A few thin bridges between islands should be noted in Fig.3.a. 

Fig.3.e shows a sample configuration at coverage $\theta $=0.5. Elongated islands have now merged to an interwoven stripe structure. Fig.3.f shows
the equivalent pair distribution taken from a configuration average. The pair distribution indicates a characteristic distance of 4 to 5 substrate
lattice spacings. \textup{ \(g_{\text{ik}}\)} values of 1.0 at \(s_{<1-10>}\)=5 to 6 and of \(1.2 \text{at} s_{<1-21>}\)=3*\(\sqrt{3}\) indicate
a weak stripe alignment towards $<$1-21$>$. We note in Fig.3.e. a similar vacancy stripe structure.

Fig.3.g shows a sample configuration at coverage $\theta $=0.9. The vacancies are forming aligned dimers, trimers, n-mers like the adatoms in Fig.3.a.

Omitting intermediate results for coverages $\theta >$0.5 has a good reason: they show vacancy structures inverse to adatom structures at (1-$\theta
$). Therefore a vacancy pair distribution { } { }

\[g_{\text{ik}}^{\text{vac}}=<\left(1-\tau _i\right) \left(1-\tau _k\right)>\left/(1-\rho )^2\right.\text{  },\text{                            }(3.1)\]

is introduced.\textup{ \(g_{\text{ik}}^{\text{vac}}\)} measures the likeliness of vacancy pairs $\{$(1-\(\tau _i\)),(1-\(\tau _k\))$\}$.

Fig.3.h shows the vacancy pair distribution taken from a configuration average at coverage $\theta $=0.9. It shows almost the same structure as the
adatom pair distribution at coverage $\theta $=0.1 in Fig.3.b, indicating a superlattice now of vacancy monomers, dimers, trimers and some 4-mers.

In summary the interaction \(u_1\) with increasing coverage leads to clusters growing on superlattice positions from mono- to n-mers. Subsequently
elongated islands are formed, merge to stripes at 0.5 ML and then the sequence is reversed with empty positions instead of occupied ones. Such changes
in adatom morphology are summarized in Tab.2. { }

\begin{doublespace}
\noindent\(\pmb{
\begin{array}{|ccccccc|}
\hline
 \text{Coverage} & \text{Form} & \text{Superlattice} & \text{Inversion} & \text{Adatoms} & \text{Vacancies} & \text{Feature} \text{Size} \\
   &   &   &   & \text{avg}. & \text{avg}. & \text{avg}. \\
 0.045 & \text{monomers} & Y &   & 1 &   & 5 \\
 0.1 & \text{dimers} & Y &   & 2 &   & 4.8 \\
 0.3 & \text{triangles}/\text{linear} & Y &   & 7 &   & 4.8 \\
 0.5 & \text{coherent} \text{stripes} &   &   &   &   & 4.6 \\
 0.9 & \text{dimers} & Y & Y &   & 2 &   \\
\hline
\end{array}
}\)
\end{doublespace}

Table 2. Changes of adatom morphology with increasing coverage, simulated with \(\text{interaction} u_1\)$\quad $

\subsection*{3.4.Influence of short range interactions, the \(U_2\) example}

Variant 1 \(U_1(s,\phi )\), Eq. (2.2) of the short range interaction was used in [4] to enable convergence of the BGY type integral equation. Compared
with Eq. (2.1) it describes an effective repulsive interaction in the short range. To show the influence of short range interactions, variant 2 \(U_2(s,\phi
)\), Eq. (2.3) is chosen less repulsive and therefore promotes adatoms to nucleate at next neighbor sites. We note in the pair distributions below
black or dark dots at next neighbor distance.

Fig.4.a shows a sample configuration with short range interaction \(U_2(s,\phi )\), Eq. (2.3) at coverage $\theta $=0.1. The superlattice consists
of many n-mers and some smaller aggregates. Fig.4.b shows the equivalent pair distribution taken from a configuration average. The black dots at
\(s_{<1-21>}\)=3*\(\sqrt{3}\) indicate a superlattice with a lattice constant of slightly above 5. The blue dots at a distance of about 10.5 in all
directions indicate a trend towards isotropy, i.e. a reduced superlattice order compared to Fig. 3.b.

Fig.4.c shows a sample configuration with short range interaction \(U_2(s,\phi )\), Eq. (2.3) at coverage $\theta $=0.3. The superlattice consists
of islands some of which have merged to elongated islands. Small bridges between islands create dog-bone-like shapes. Fig.4.d shows the equivalent
pair distribution taken from a configuration average. The blue dots indicate an isotropic ring structure with a characteristic distance of nearly
6, the \textup{ \(g_{\text{ik}}\)} values of 1.2 at \(s_{<1-10>}\)=5 and of \(1.35 \text{at} s_{<1-21>}\)=3*\(\sqrt{3}\), however, indicate a weak
island alignment towards $<$1-21$>$. { }

Fig.4.e shows a sample configuration with short range interaction \(U_2(s,\phi )\), Eq. (2.3) at coverage $\theta $=0.5. The islands of lower coverages
have now merged to an interwoven but incoherent stripe structure with an average stripe broadness of nearly 3. Fig.4.f shows the equivalent pair
distribution taken from a configuration average. The blue dots again pretend an isotropic ring structure with a characteristic distance of 6, the
\textup{ \(g_{\text{ik}}\)} values of 1.1 at \(s_{<1-10>}\)=5 to 6 and of \(1.2 \text{at} s_{<1-21>}\)=3*\(\sqrt{3}\), however, indicate a weak stripe
alignment towards $<$1-21$>$. Compared with Fig. 3.c - with short range interaction \(U_1\) at $\theta $=0.5 - the stripes are a bit thicker, slightly
less coherent and their distance is about one lattice constant larger.

\subsection*{3.5.Influence of short range interactions, the \(U_3\) example}

Variant 3 \(U_3(s,\phi )\), Eq. (2.4) is more attractive and therefore strongly promotes adatoms to nucleate at next neighbor sites. We note in the
pair distributions black dots at 1 lattice spacing in the $<$1-10$>$ direction reflecting strong population of next neighbor sites.

Fig.5.a shows a sample configuration with short range interaction \(U_3(s,\phi )\), Eq. (2.4) at coverage $\theta $=0.1. The cluster structure consists
of a variety of sizes from monomers to n-mers. Fig.5.b shows the equivalent pair distribution taken from a configuration average. A reduced alignment
of clusters to the substrate crystal directions is visible compared to Figs.3.b and 4.b. 

Fig.5.c shows a sample configuration with short range interaction \(U_3(s,\phi )\), Eq. (2.4) at coverage $\theta $=0.3. The cluster structure consists
of larger islands some of which have merged to elongated islands with dog-bone-like shapes. Fig.5.d shows the equivalent pair distribution taken
from a configuration average. The blue dots pretend an isotropic ring with a characteristic distance of 6, the \textup{ \(g_{\text{ik}}\)} values
of 1.1 at \(s_{<1-10>}\)=5 to 6 and of \(1.3 \text{at} s_{<1-21>}\)=3*\(\sqrt{3}\), however, indicate a weak island alignment towards $<$1-21$>$.

Fig.5.e shows a sample configuration with short range interaction \(U_3(s,\phi )\), Eq. (2.4) at coverage $\theta $=0.5. The islands of lower coverages
have merged to an interwoven stripe structure with an average stripe broadness of more than 3. Fig.5.f shows the equivalent pair distribution taken
from a configuration average. Within the range of blue dots \textup{ \(g_{\text{ik}}\)} values of 1.05 at \(s_{<1-10>}\)=5 to 6 and of \(1.1 \text{at}
s_{<1-21>}\)=3*\(\sqrt{3}\)indicate very weak alignment towards $<$1-21$>$. The characteristic distance is between 5 and 6. Compared with Fig. 4.e
- with short range interaction \(U_2\) at $\theta $=0.5 - the stripes look similar, but if Fig.5.f is compared with Fig.4.f we note a tendency towards
a reduced order.

In summary the variants with more attractive short range interactions lead to less ordered superlattices consisting of more adatoms with a greater
superlattice constant and thicker, more coherent stripes.

\section*{4.Discussion}

In this section the model assumptions are reviewed, the model results are summarized and compared with experiments. The section closes with a discussion
of open points.

\subsection*{4.1.Model assumptions}

Assumptions and approximations used for this model have been discussed in [4] in detail. The most relevant approximation is an elastic continuum
model for the substrates instead of a lattice model, known to be inadequate for describing short range effects. The elastic continuum model predicts
a \(s^{-3}\) repulsion on the long range, a repulsive wall near 2.3 lattice distances and a strong attractive well at next neighbor distances.\\
The assumption of a perfectly flat surface excludes the effects of steps, known for their active role in nucleation and growth, partly due to strain
in their neighborhood. \\
A further key assumption is thermal equilibrium for the adatom configurations, i.e. neglecting of kinetic effects.\\
Anisotropic stress generated by stretching adatom bonds is not covered.\\
Further assumptions cover the short range interactions used. Replacement of the deep attractive potential well by either a cap of about 5 units (in
fact describing a strong repulsion of next neighboring adatoms) or by a cap at potential zero (in fact describing a weak repulsion of next neighboring
adatoms) or by a cap at potential -5 units (describing a stronger attraction of next neighbors) is a method to indicate the effects of short range
interactions while preserving the merits of a theory with medium range focus. 

\subsection*{4.2.Comparison with previous off-grid simulations}

The adatom configurations with interaction \(U_1\) at $\theta $=0.045 resulting from an off-grid algorithm in [4] could be repeated with an on-grid
algorithm. Both results and the derived pair distributions are consistent with the solution of a Born-Green-Ivon type integral equation describing
the adatom pair distribution from Statistical Mechanics principles [7]. So the present simulations can be rated as a high coverage extension of previous
results. 

\subsection*{4.3.Adatom distribution on Cu $<$111$>$}

The results of sections 3.1 to 3.3 draw the picture of a substrate aligned hexagonal packed superlattice of adatoms or clusters at coverages up to
$\theta \approx $0.3 ML and stripes around coverages $\theta $=0.5 ML. The simulations predict a populated/unpopulated symmetry $\theta $/(1-$\theta
$). This is explained by minimization of the total configuration energy by forming adatom clusters or stripes or vacancies around 5 lattice distances
apart. Substrate strain generated by adatoms stressing the surface is allowed to release within the unpopulated sites. 

\subsection*{4.4.The role of short range interactions}

The results of sections 3.4 and 3.5 indicate a strong role of short range interactions. The trends when increasing the short range attraction are\\
- larger clusters\\
- triangular instead of linear clusters (reflected by an increasing next neighbor pair distribution) \\
- more connected clusters\\
- less influence of the substrate on the superlattice and stripe directions, i.e. more isotropic configurations\\
- slightly increased feature size. { }

\subsection*{4.5.Comparison with experiments}

Strain mediated superlattices on Cu$<$111$>$ in the coverage region up to $\theta $=0.045 ML were already discussed and compared with experiments
[10,11] in [4]. The agreement was good enough to propose strain mediated interactions as an alternative to the one discussed in [10,11]. The current
investigation was triggered by an unknown referee of [4]. He raised the point how other manifestations of elastic interactions on surfaces, especially
stress domains [12], are related to adatom superstructures.

Stress domains are ordered patterns of less dense and more dense adatom areas, for example adatom gas areas and monolayer areas forming spontaneously.
They minimize surface energy by balancing short-range attractive with long-range repulsive interaction. Substrate strain created by surface stress
in the more dense areas is allowed to relax in the less dense areas. The domains reflect the elastic anisotropy of the substrate. 

Observations at the Pb/Cu$<$111$>$ system [5] can be characterized by \\
- ordered but mobile circular droplets (containing thousands of adatoms) at low coverages\\
- stripes at medium coverage with a long range order improving when reducing temperature\\
- ordered inverse droplets at high coverages approaching a monolayer (as predicted earlier, see references in [5]).\\
The periodicity of patterns is in the 100 nm range, decreasing with increasing temperature. The observed temperature range is 623 K to 673 K. The
order of droplets can - from a first glance - be interpreted as a superlattice type. 

The sequence of island superlattices, domain patterns and inverse droplets on Pb/Cu$<$111$>$ with increasing coverage observed in [5] would serve
as a striking experimental evidence of the theory and the simulation results presented in section 3.4 { }if the length scales and the temperature
would be the same. Unfortunately [5] describes a high temperature experiment with adatom clusters of thousands of adatoms while the experiments showing
superlattice effects on a few lattice constant scale have been performed in the 10K region [10,11]. So the question arises if the present Molecular
Dynamics simulation could be extended to handle clusters and structures of hundreds or thousands of adatoms. Unfortunately this would be far beyond
the resources of a { }PC, so we must rely on scaling arguments to argue the same driver - elastic interactions - for both phenomena, adatom superlattices
and stress domain patterns:

Following the cluster section of [8] we argue a simple superposition ansatz for the elastic interactions: two clusters of \(n_1\) and \(n_2\) adatoms
create \(n_1\)*\(n_2\) times the elastic inter cluster energy of two single adatoms. This ansatz, of course, is a strong restriction not considering
e.g. short range adatom-adatom interactions and lattice mismatch effects.

Two n=\(10^3\) clusters would create an interaction \(U^{\text{Cluster}}\) \(10^6\) times \(U\). The { }temperature range for the stress domain experiments
at 650 K is a factor\(10^2\) higher than the regime of single adatom effects, so the scaled cluster interaction \(u^{\text{Cluster}}\)=\(U^{\text{Cluster}}\)/\(k_B\)T
would be \(10^4\) times higher. The typical length \textup{ \(s^{\text{Cluster}}\)} according to Eq. (2.9) would then be \(10^{4/3}\)$\approx $21
times higher than \textup{ \(s\)}. The coverage \(\theta ^{\text{Cluster}}\) would be reduced by a factor of about 464 (noting that the coverage
of adatoms and of clusters have different meaning).

We summarize that length scales may differ from 5 substrate lattice constants to 100 lattice constants in the example above, but the elastic interaction
mechanism is the same. { } 

It would be a big surprise if such simple scaling arguments could explain the physics of stress domains more than qualitatively. In fact the measurement
in [15] shows a decrease in domain feature size from 140 nm to 40 nm when the temperature rises from 590 K to 650 K, far beyond the above scaling
effects. The authors explain such decrease in stripe periodicity with the change in domain-boundary free energy caused by thermally broken Pb-Pb
bonds. Thus the effects of short range interactions override the effects of elastic medium range interactions under certain conditions. This is not
in contradiction to the scaling arguments above since Eq. (2.9) is valid only for a \textup{ \(s^{-3}\)} type interaction which does not include
short range effects. { }

The influence of substrate anisotropy as observed in [13] is reflected in the present simulations and the pair distributions derived. The dominating
stripe orientation at coverage $\theta $=0.5 ML reported is $<$-1-12$>$, the model results show a weak stripe orientation in the same directions.
{ }.

The assumed equilibrium conditions are confirmed by experiments with reversible shape transitions (droplet to elongated islands with dog-bone-like
shapes) during heating cycles [14]. An increase in temperature changes the shapes the same way as an increasing coverage does, again in line width
the scaling arguments outlined above (same shape of \textup{ \(g\left(s,u_0,\theta \right)\)} if \textup{ \(u_0\)} reduced or $\theta $ increased).
Both experimental results can be seen as a hint for the validity of the assumptions and conclusions.

The lack of details reported makes the comparison of further experiments with the present model similarly difficult. Two further examples should
show its ability and its limitations: { } 

N on Cu$<$111$>$ forms elongated islands, stable at room temperature, aligned in 3 equivalent crystal directions. The islands show a characteristic
distance of about 10 nm and are often colliding [16]. Since coverage and crystal directions are not reported, a comparison with stripes as calculated
from the model is incomplete, but similarities with Figs. 3.e, 4.e, 5.e should encourage further research. { }

Co on Cu$<$111$>$ acts as a nanoisland reference system with a well documented strain mediated morphologies [17, 18]. They are different from the
ones found in this calculations due to their tendency of bi-layer growth even at moderate coverages. Triangular bi-layer islands show lateral displacements
of the Co-Co bond lengths (measured by the surface state electron energy) dependent on their positions within the islands, associated with lateral
strain.

The following picture for strain mediated morphologies is concluded:\\
Stripe morphologies correlate with repulsive short range interaction while attractive short range interactions destabilize stripes and - via multilayer
growth - lead to islands. Islands create strain and interact via strain and their shape minimizes elastic energy. { } { }

\subsection*{4.6.Open questions}

Clusters arranged in superlattices and stress domain patterns on Cu$<$111$>$ in the temperature range of about 10K with a characteristic length of
about 5 to 6 lattice constants hopefully may be found in existing material. More experimental material is needed to determine size and nature of
short range interactions and also the orientation of cluster/ stripe structures relative to the substrate crystal directions.

First principles methods (like density-functional-theory) need to be applied for estimating the stress parameters.

A further question is how the theory successfully describing mesoscopic stress patterns [12] can be utilized to better understand the microscopic
effects discussed in the present analysis: When domain boundary effects play an important role in the mesoscopic range, the effects of short range
interactions in the microscopic range should be similarly significant. { } 

The morphology of adatoms on other surfaces is an equally interesting topic, the much stronger effects of elastic anisotropy on $<$001$>$ surfaces
of many materials is expected to lead to further insight. Increasing the accuracy of the simulation by increasing the diameter of the simulation
area on much more powerful computers may also lead to additional insights. 

Bi-layer effects would extend the scope of the model but require some basic work. It should be noted that long range magnetic interactions should
also be considered in the Co case. 

The restrictions of the present isotropic stress model motivated an extension of the model [19]: Dimers are supposed to create anisotropic stress
by stretching their bond. Such stress creates other types of elastic interactions including lattice mismatch.

\section*{7.Summary}

Substrate strain mediated adatom configurations have been simulated for Cu$>$111$>$ surfaces for three short range interaction types. The adatom
coverages range up to nearly a monolayer. Pair distributions have been derived to prove morphologies from superlattices of single adatoms and clusters
to ordered stress domain patterns. Higher coverages beyond 0.5 monolayers show vacancy structures just inverted. The short range interaction shows
a significant influence on the cluster size within the superlattices. Substrate elastic anisotropy influences the superlattice orientation with respect
to the substrate crystal directions.

Experiments showing similar structures have been compared with the model. For low temperatures superlattices of single adatoms have been found while
for increased temperatures ordered islands and stripes of adatoms have been reported. There is some evidence of elastic interactions being the common
cause but a final conclusion on the validity of the theory remains open at this point in time.

\section*{Erratum}

In the course of recent calculations a code fault affecting previous results [4,7] was detected: The \(\cos (p \pi /2)\) term in Eq. (2.1) was omitted
in the code there. Therefore the p=6 interaction terms on $<$111$>$ surfaces had a sign error. As a consequence the results have to be rotated by
30${}^{\circ}$. In the current version of this paper the fault has been corrected. { }The author apologizes for any inconvenience. 

\section*{References}

[1] H.Brune, Creating Metal Nanostructures at Metal Surfaces Using Growth Kinetics, in: Handbook of Surface Science Vol.3 (E.Hasselbrink and B.I.Lundqvist
ed.), Elsevier, Amsterdam (2008) 

[2] H.Ibach, Surf.Sci.Rep. 29, 195 (1997)

[3] T.L.Einstein, Interactions between Adsorbate Particles, in: Physical Structure of Solid Surfaces (W.N.Unertl ed.), Elsevier, Amsterdam (1996)

[4] W.Kappus, Surf. Sci. 609, 30 (2013)

[5] R.Plass, J.A.Last, N.C.Bartelt, G.L.Kellogg, Nature 412, 875 (2001)

[6] L.Proville, Phys. Rev. B 64, 165406 (2001)

[7] W.Kappus, Surf. Sci. 606, 1842 (2012) 

[8] W.Kappus, Z.Physik B 29, 239 { }(1978) 

[9] A.G.Every, A.K.McCurdy: Table 3. Cubic system. Elements. D.F.Nelson(ed.), SpringerMaterials- The Landolt-B{\" o}rnstein Database 

[10] J.Repp, F.Moresco, G.Meyer, K.H.Rieder, P.Hyldgaard and M.Persson, Phys. Rev. Lett. 85, 2981 (2000).

[11] F.Silly, M.Pivetta, M.Ternes, F.Patthey, J.P.Pelz, W.D.Schneider, New J. of Phys. 6, 16 (2004)

[12] O.L.Alerhand, D.Vanderbilt, R.D.Meade, J.D.Joannopoulos, Phys. Rev. Lett. 61, 1973 (1988)

[13] F.Leonard, N.C.Bartelt, G.L.Kellogg, Phys. Rev. B 71, 045416 (2005)

[14] R.van Gastel, N.C.Bartelt, G.L.Kellogg, Phys. Rev. Lett. 96, 036106 (2006)

[15] R.van Gastel, N.C.Bartelt, P.J.Feibelman, F.L{\' e}onard, and G.L.Kellogg, Phys. Rev. B 70, 245413 (2004)

[16] F.M.Leibsle, Surf. Sci. 514, 33 (2002)

[17] M.V.Rastei, B.Heinrich, L.Limot, P.A.Ignatiev, V.S.Stepanyuk, P. Bruno, J.P.Bucher, Phys. Rev. Lett. 99, 246102 (2007)

[18] N.N.Negulyaev, V.S.Stepanyuk, P. Bruno, L.Diekh{\" o}ner, P.Wahl, K.Kern, Phys. Rev. B 77, 125437 (2008)

[19] W.Kappus, http://arxiv.org/abs/1301.3643

\section*{Acknowledgement}

Many thanks to the unknown referee of [4] for directing the author{'}s interest to stripes. 

\section*{Appendix}

\includegraphics{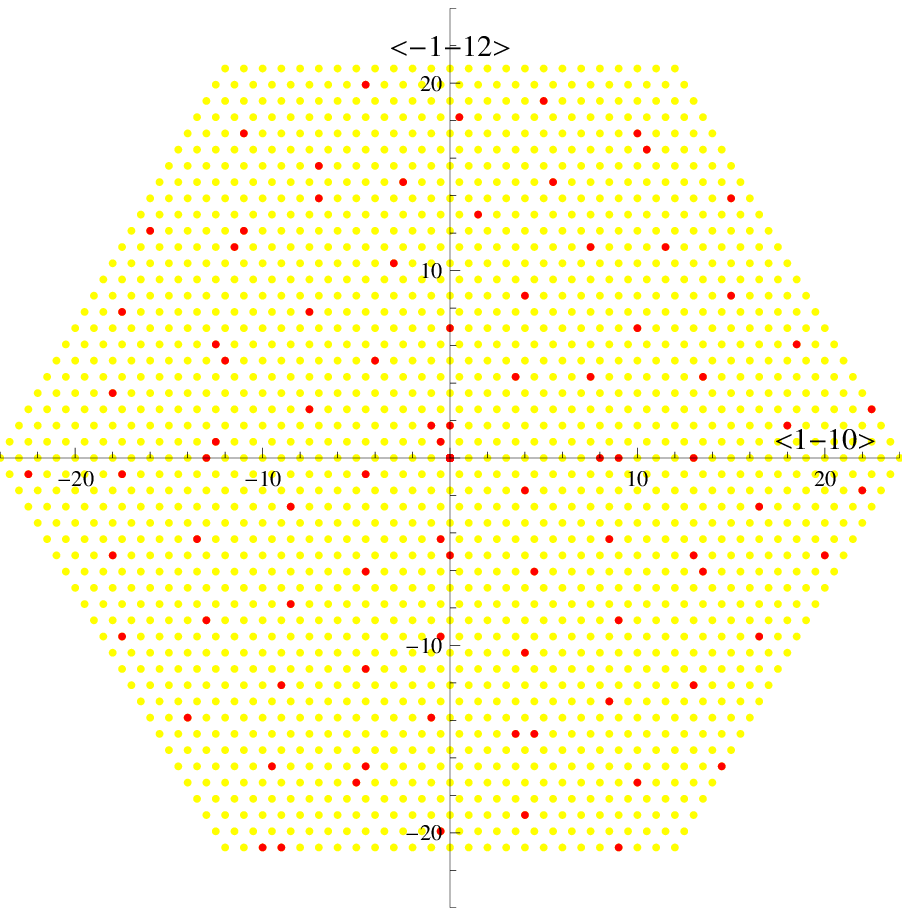}

Fig. 1.a shows a sample configuration with reference short range interaction \(U_1(s,\phi )\) (2.2) and a coverage $\theta $=0.045.

\includegraphics{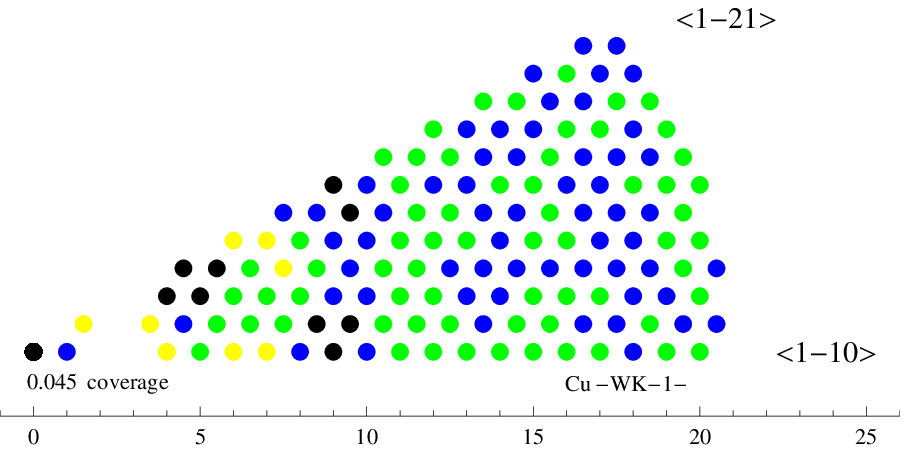}

Fig. 1.b shows an average adatom pair distribution in a 30 degree sector with reference short range interaction \(U_1(s,\phi )\) (2.2) and a coverage
$\theta $=0.045. The differently colored dots represent different values of the pair distribution (darker colors represent higher values).

\includegraphics{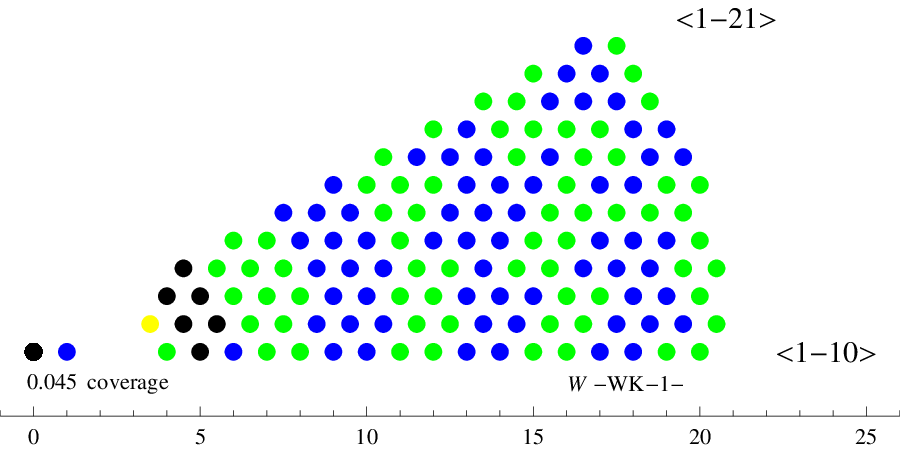}

Fig.2 shows an average pair distribution for an isotropic substrate (tungsten) at coverage { }$\theta $=0.045 in a 30 degree sector.

\includegraphics{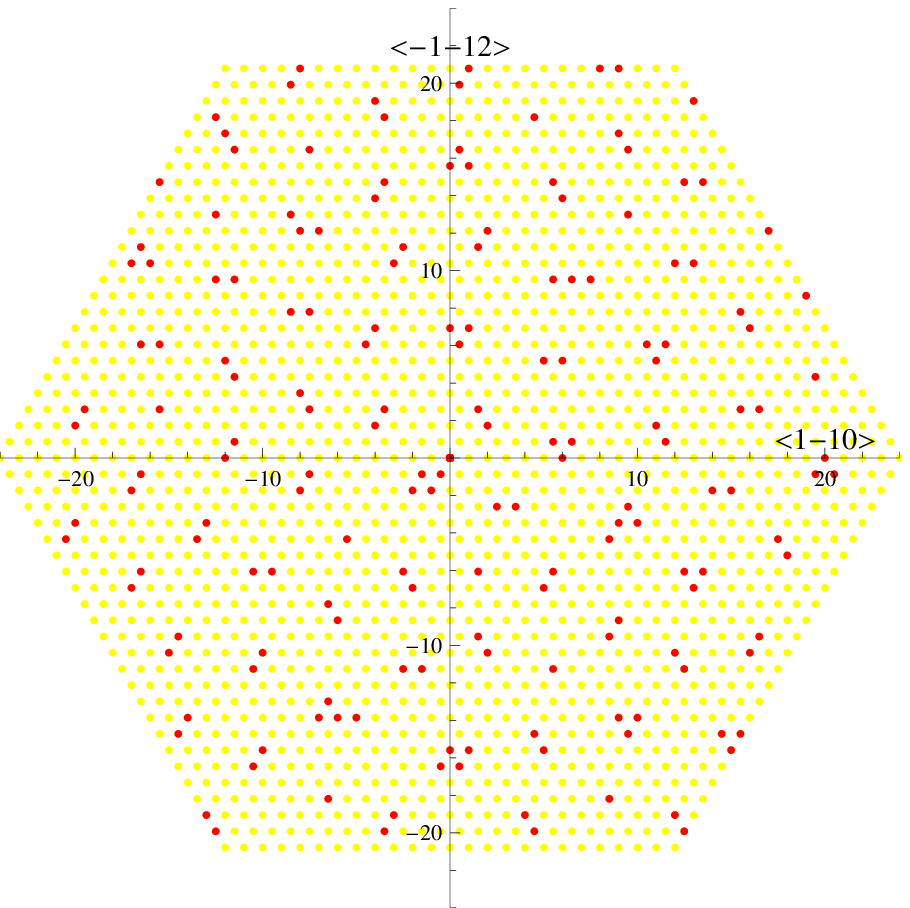}

Fig.3.a shows a sample configuration with short range interaction \(U_1(s,\phi )\) (2.2) at coverage $\theta $=0.1.

\includegraphics{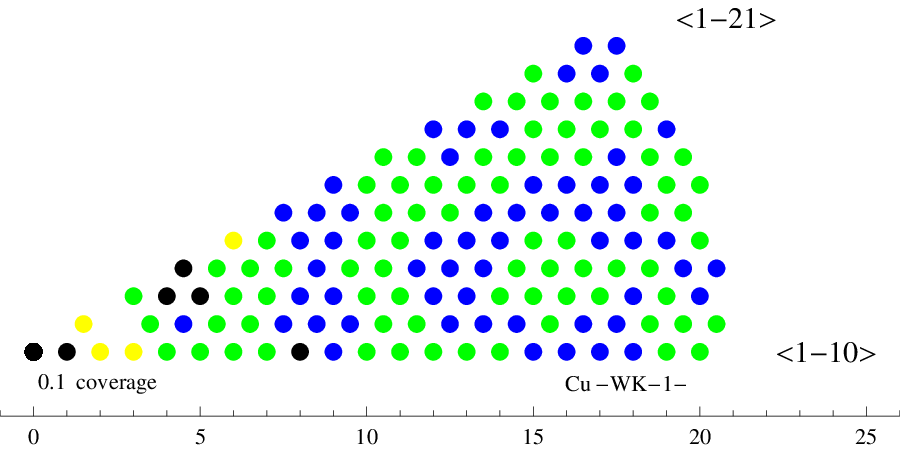}

Fig.3.b shows an average pair distribution with short range interaction \(U_1(s,\phi )\) (2.2) at coverage $\theta $=0.1.

\includegraphics{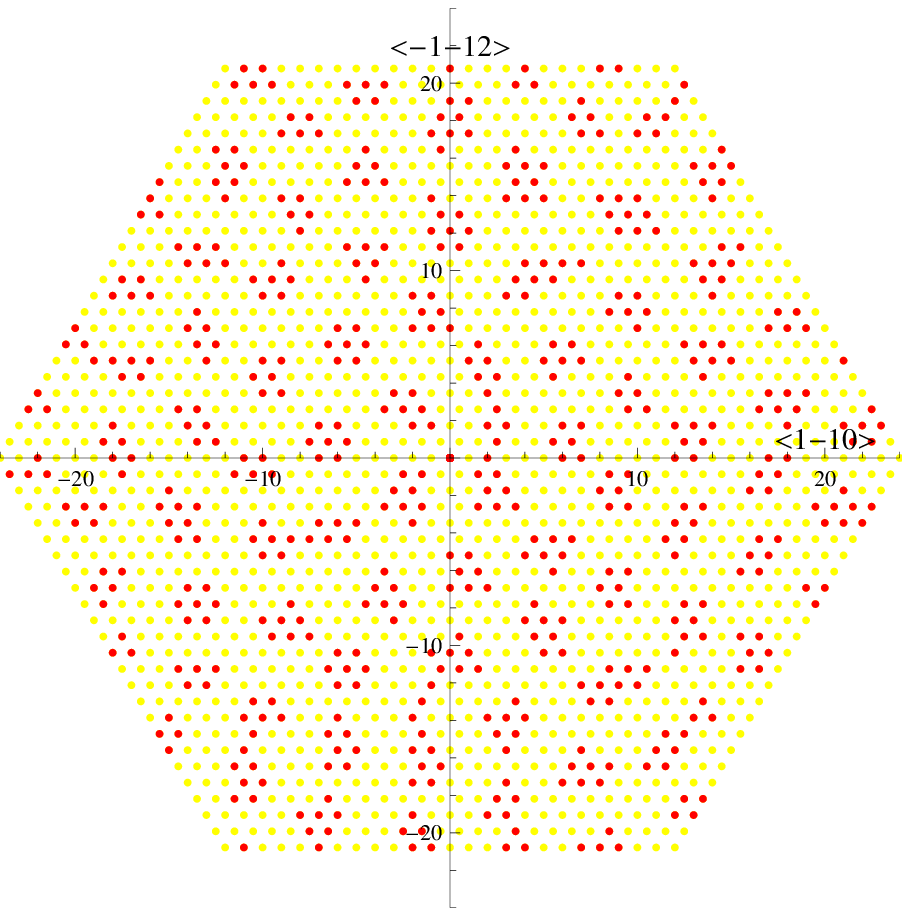}

Fig.3.c shows a sample configuration with short range interaction \(U_1(s,\phi )\) (2.2) at coverage $\theta $=0.3.

\includegraphics{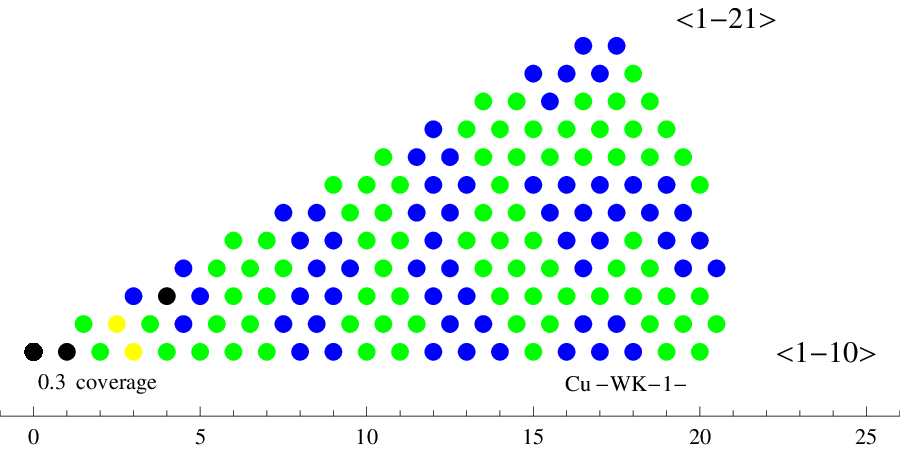}

Fig.3.d shows an average pair distribution with short range interaction \(U_1(s,\phi )\) (2.2) at coverage $\theta $=0.3.

\includegraphics{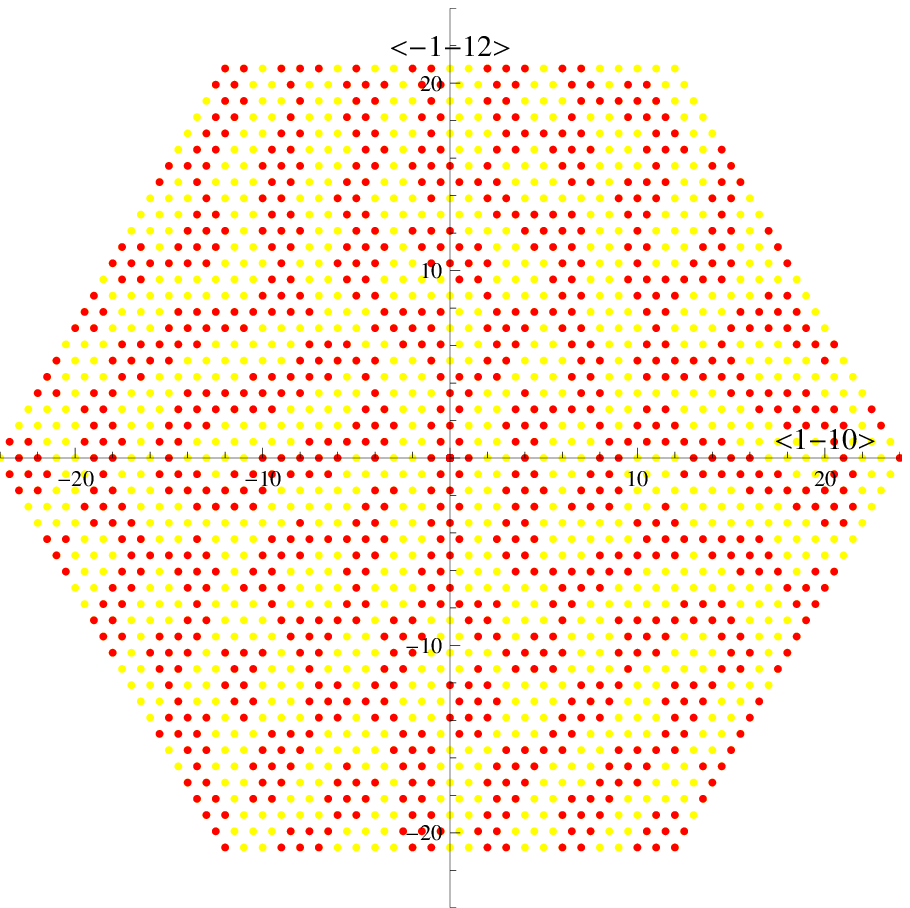}

Fig.3.e shows a sample configuration with short range interaction \(U_1(s,\phi )\) (2.2) at coverage $\theta $=0.5.

\includegraphics{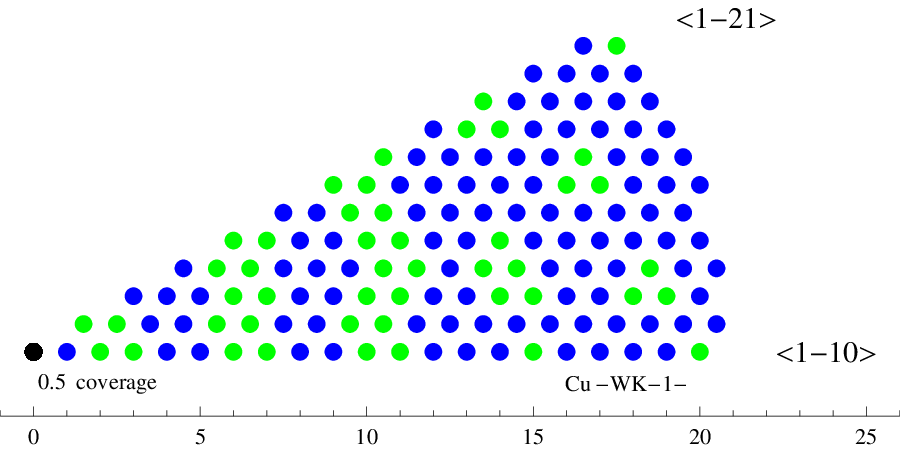}

Fig.3.f shows an average pair distribution with short range interaction \(U_1(s,\phi )\) (2.2) at coverage $\theta $=0.5.

\includegraphics{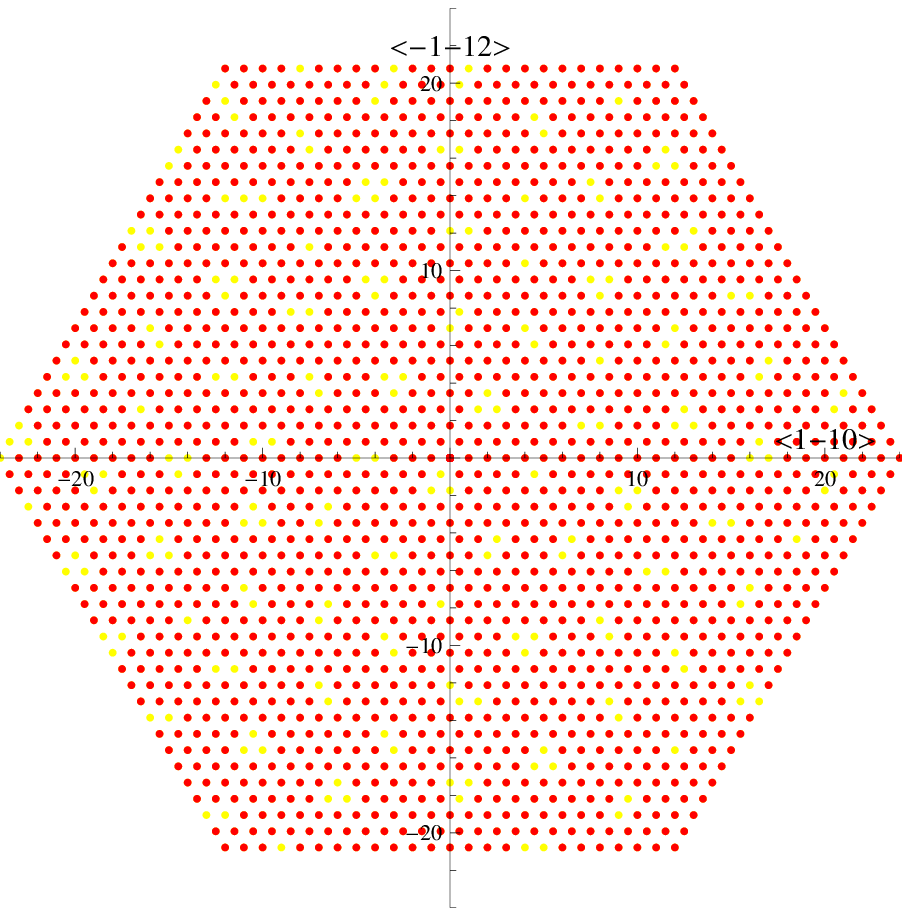}

Fig.3.g shows a sample configuration with short range interaction \(U_1(s,\phi )\) (2.2) at coverage $\theta $=0.9.

\includegraphics{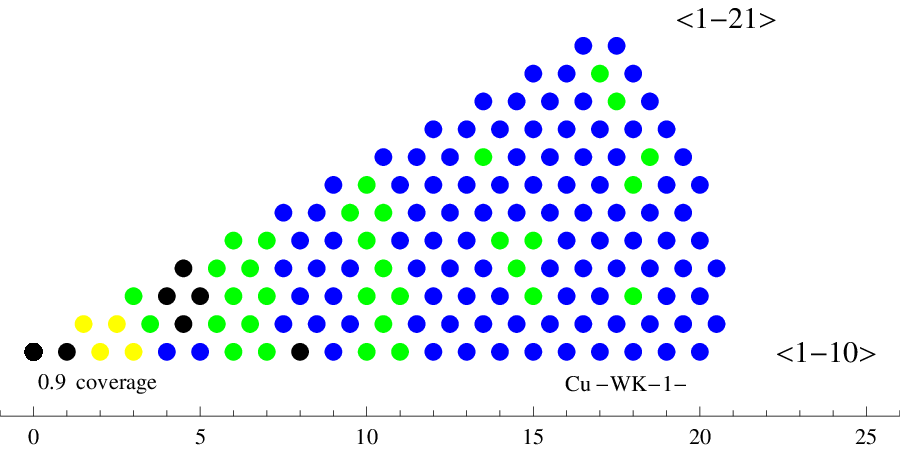}

Fig.3.h shows an average vacancy pair distribution (3.1) with short range interaction \(U_1(s,\phi )\) (2.2) at coverage $\theta $=0.9.

\includegraphics{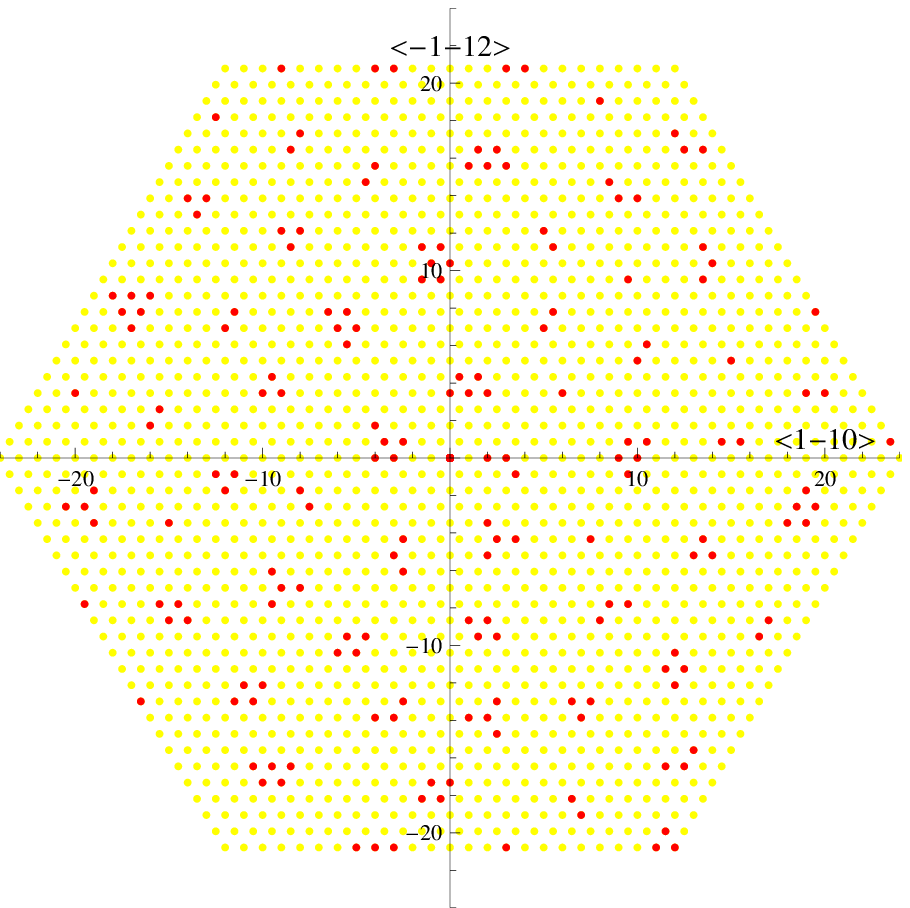}

Fig.4.a shows a sample configuration with short range interaction \(U_2(s,\phi )\) (2.3) at coverage $\theta $=0.1. 

\includegraphics{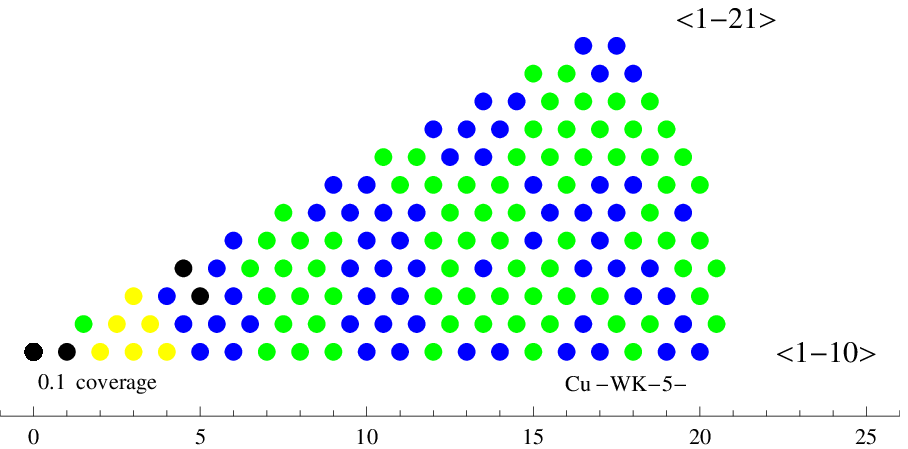}

Fig.4.b shows an average pair distribution with short range interaction \(U_2(s,\phi )\) (2.3) at coverage $\theta $=0.1.

\includegraphics{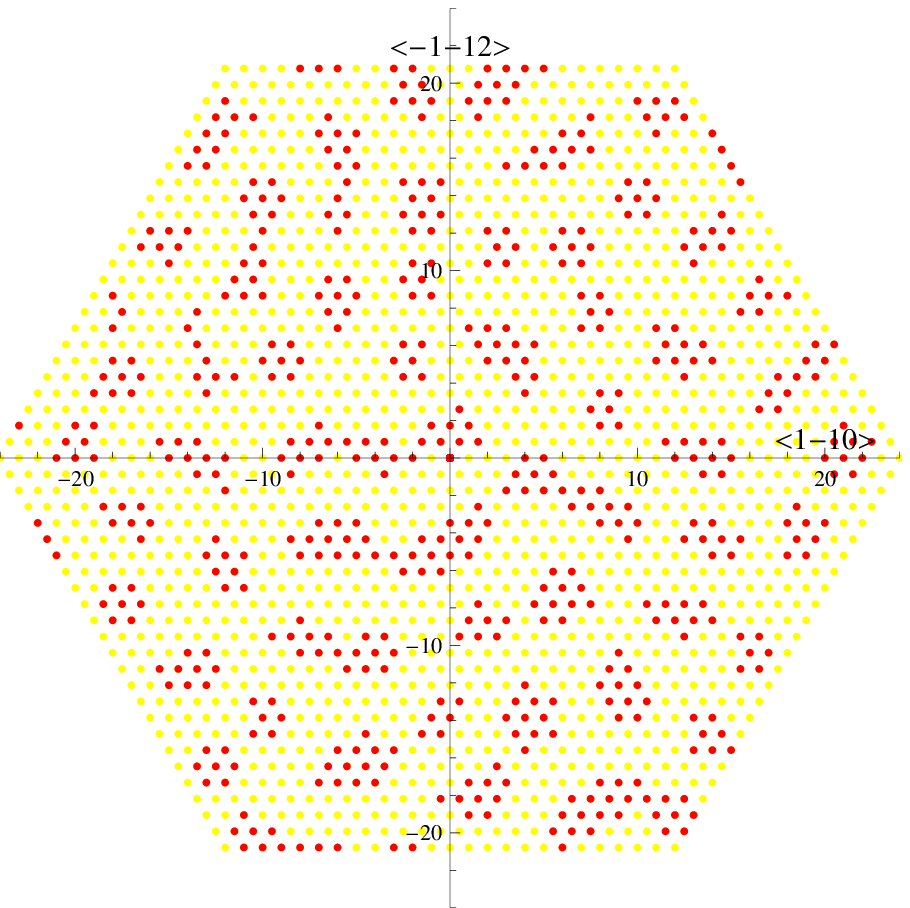}

Fig.4.c shows a sample configuration with short range interaction \(U_2(s,\phi )\) (2.3) at coverage { }$\theta $=0.3.

\includegraphics{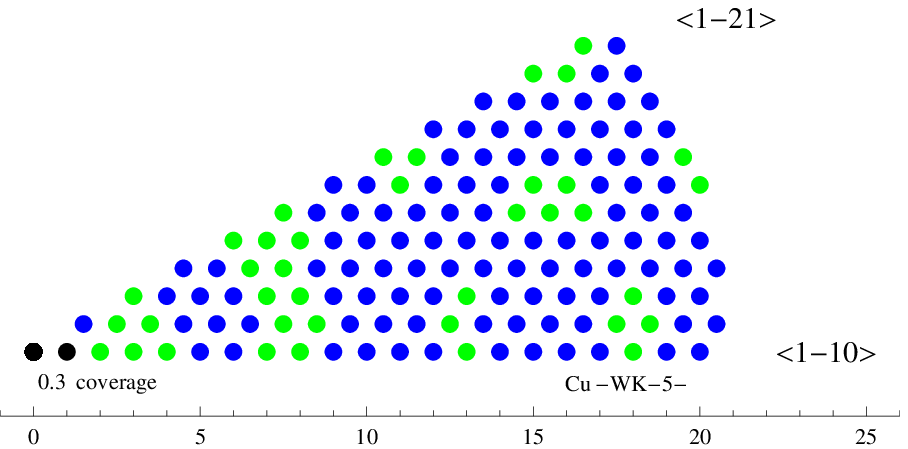}

Fig.4.d shows an average pair distribution with short range interaction \(U_2(s,\phi )\) (2.3) at coverage { }$\theta $=0.3.

\includegraphics{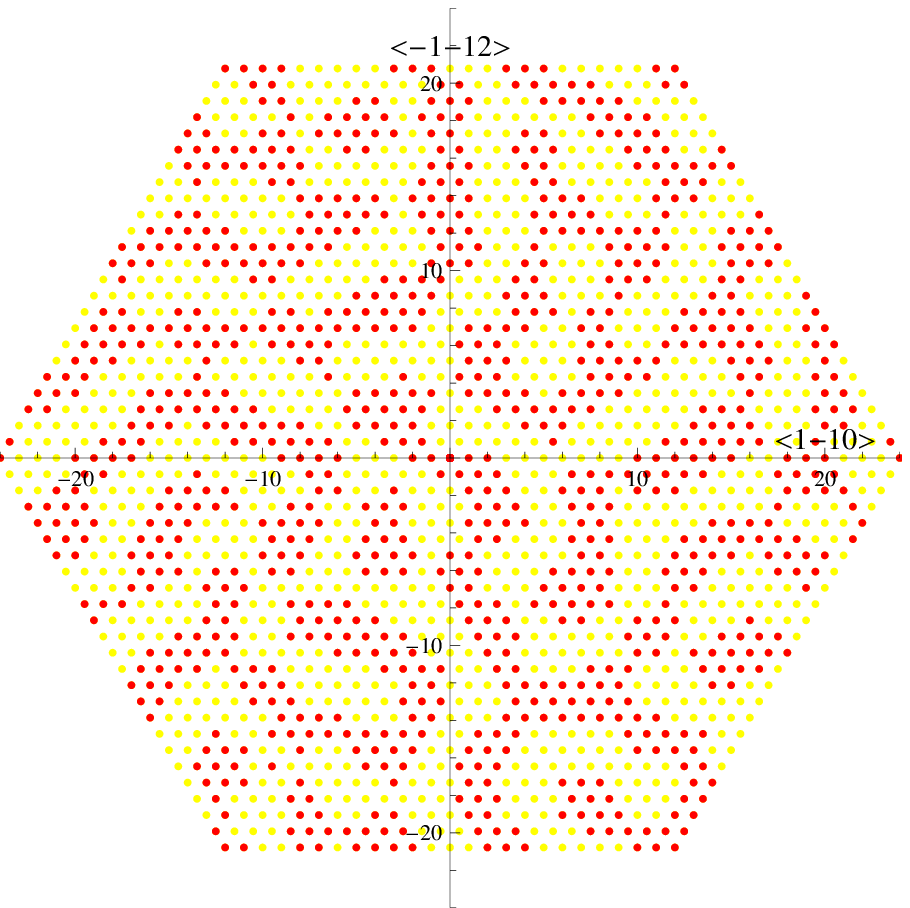}

Fig.4.e shows a sample configuration with short range interaction \(U_2(s,\phi )\) (2.3) at coverage $\theta $=0.5. 

\includegraphics{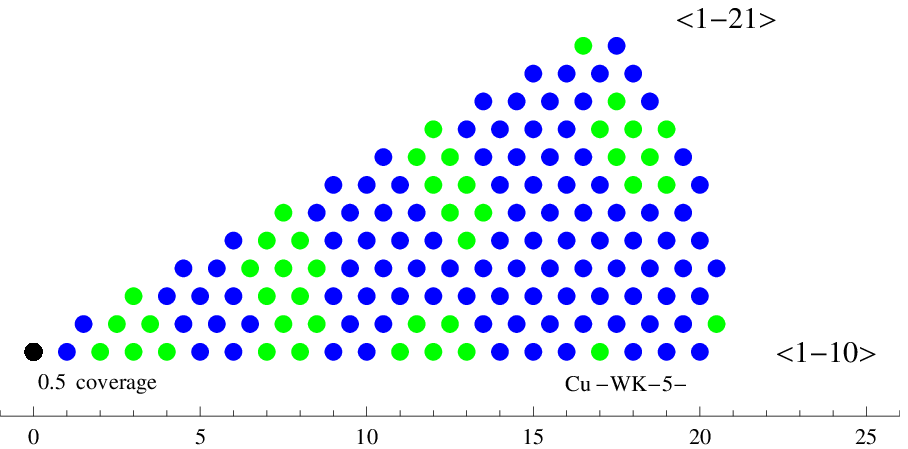}

Fig.4.f shows an average pair distribution with short range interaction \(U_2(s,\phi )\) (2.3) at coverage $\theta $=0.5.

\includegraphics{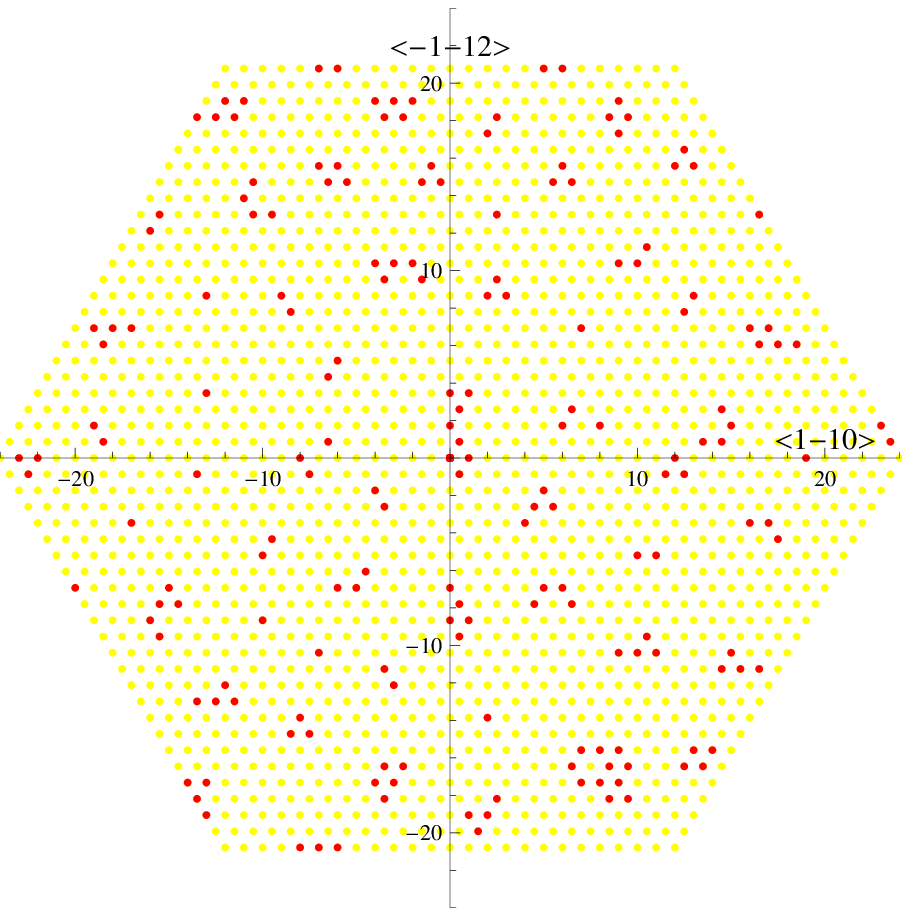}

Fig.5.a shows a sample configuration with short range interaction \(U_3(s,\phi )\) (2.3) at coverage $\theta $=0.1. 

\includegraphics{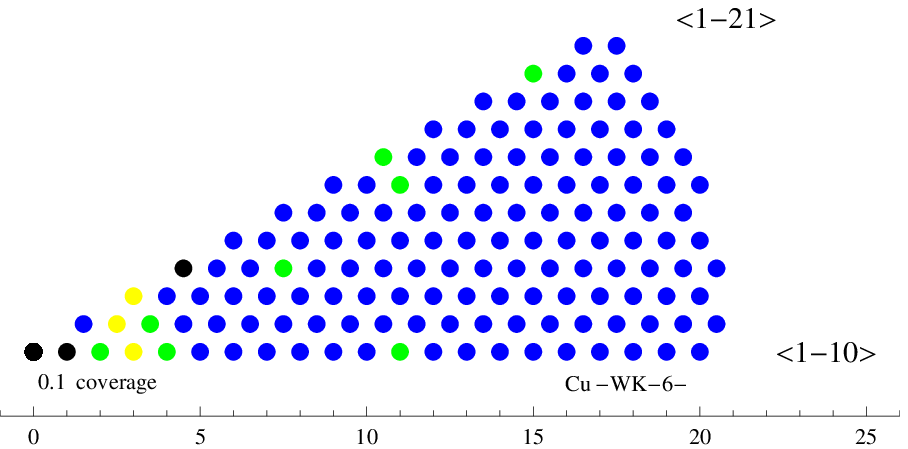}

Fig.5.b shows an average pair distribution with short range interaction \(U_3(s,\phi )\) (2.3) at coverage $\theta $=0.1.

\includegraphics{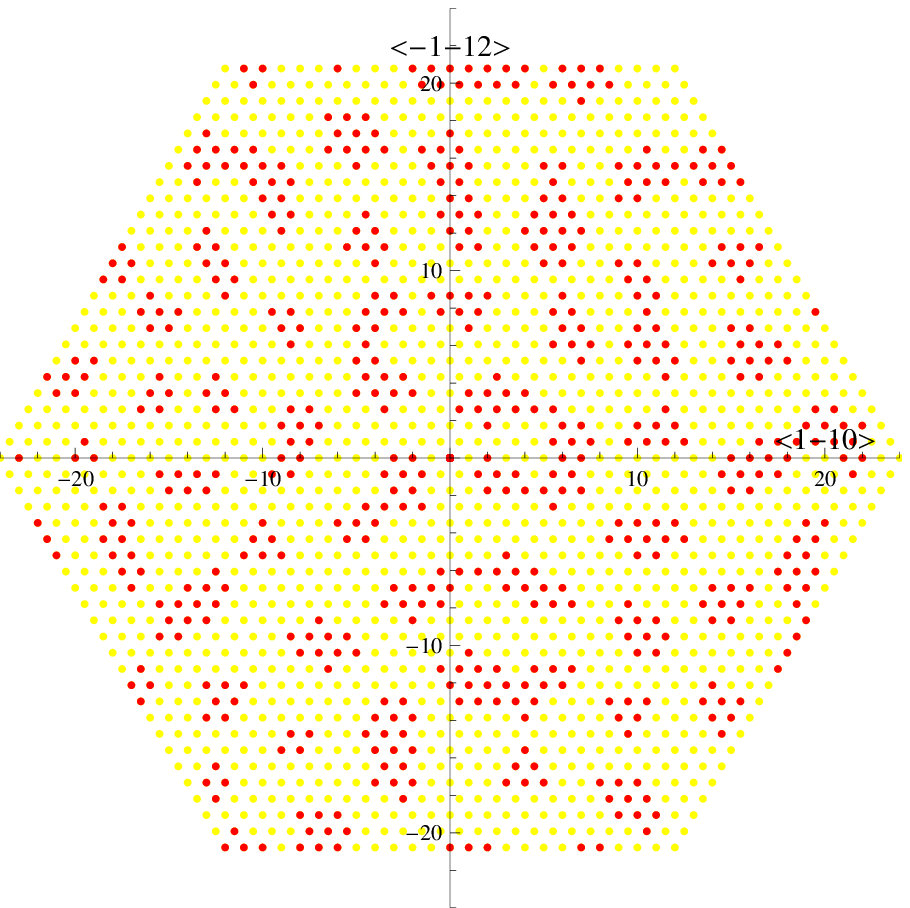}

Fig.5.c shows a sample configuration with short range interaction \(U_3(s,\phi )\) (2.3) at coverage { }$\theta $=0.3.

\includegraphics{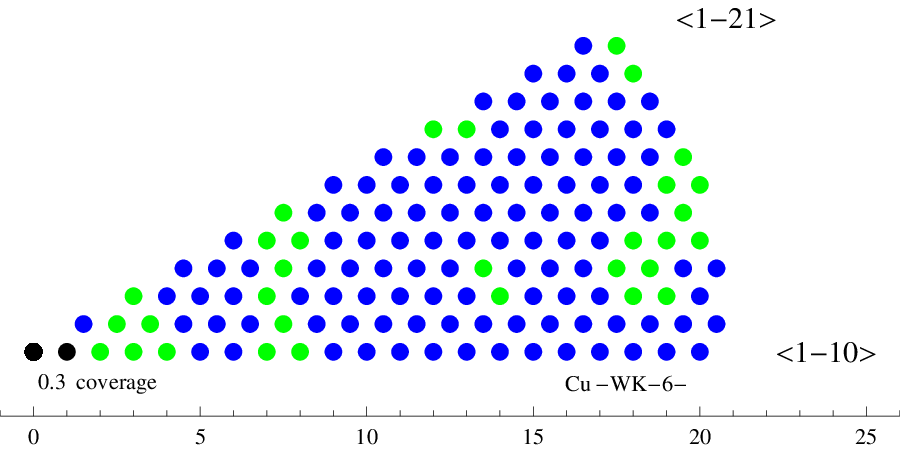}

Fig.5.d shows an average pair distribution with short range interaction \(U_3(s,\phi )\) (2.3) at coverage { }$\theta $=0.3.

\includegraphics{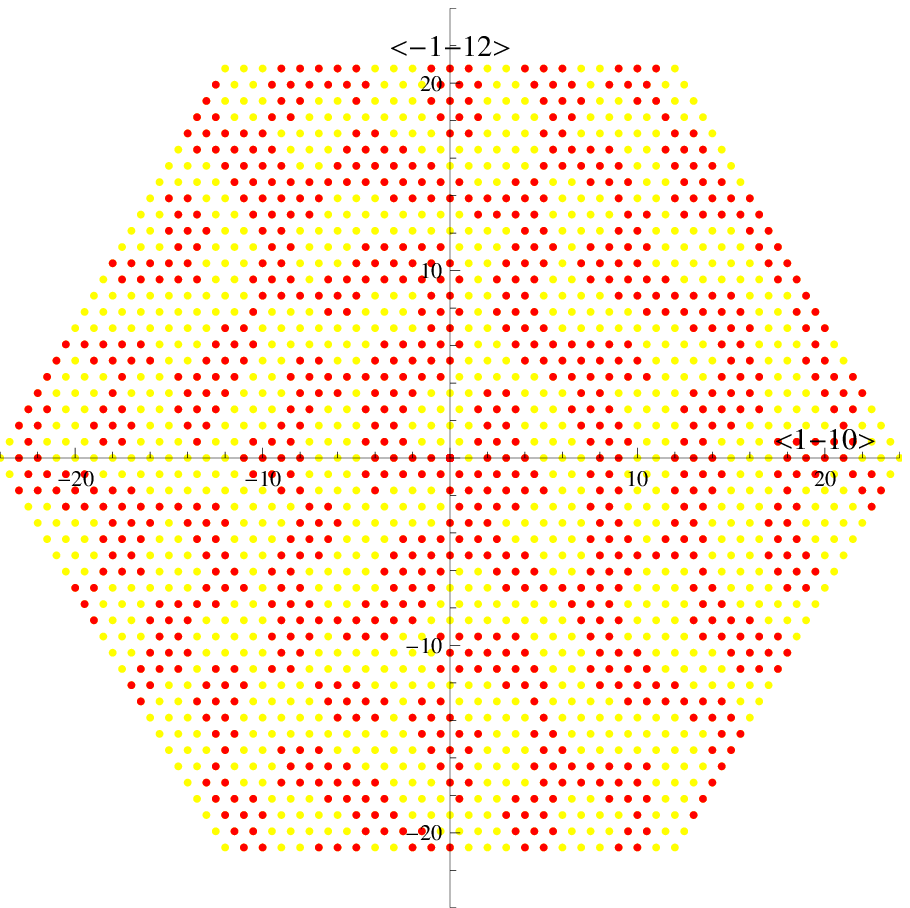}

Fig.5.e shows a sample configuration with short range interaction \(U_3(s,\phi )\) (2.3) at coverage $\theta $=0.5. 

\includegraphics{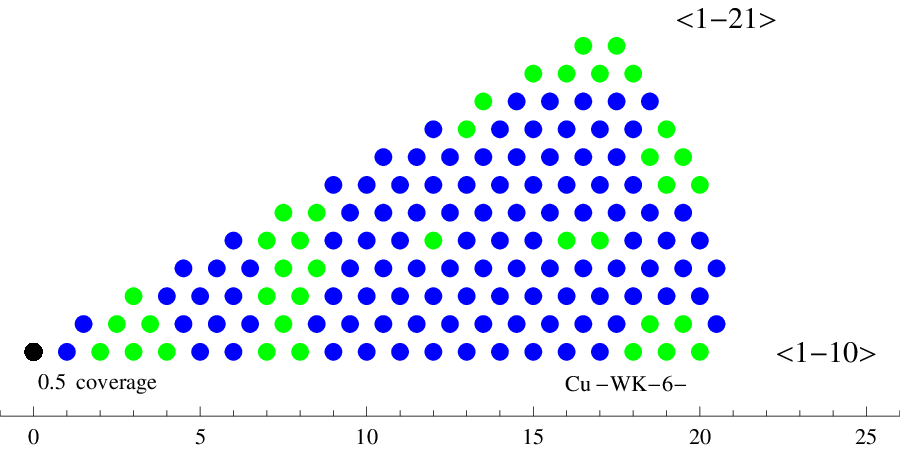}

Fig.5.f shows an average pair distribution with short range interaction \(U_3(s,\phi )\) (2.3) at coverage $\theta $=0.5.

$\copyright $ Wolfgang Kappus 2012\\

\end{document}